\documentclass[epj]{svjour}

\usepackage{graphicx}

\begin{document}

\title{Numerical analysis of reversible
       A $+$ B $\leftrightarrow$ C reaction-diffusion systems}

\author{Zbigniew Koza\thanks{e-mail: {\tt zkoza@ift.uni.wroc.pl}}}

\institute{Institute of Theoretical Physics, University of Wroc{\l}aw, plac
Maxa Borna 9, PL-50204 Wroc{\l}aw, Poland}

%\date{\today}

\abstract{We develop an effective numerical method of studying large-time
properties of reversible reaction-diffusion systems of type A $+$ B
$\leftrightarrow$ C with initially separated reactants. Using it we find
that there are three types of asymptotic reaction zones. In particular we
show that the reaction rate can be locally negative and concentrations of
species A and B can be nonmonotonic functions of the space coordinate $x$,
locally significantly exceeding their initial values.}


\maketitle

%%%%%%%%%%%%%%%%%%%%%%%%%
%%%  I. INTRODUCTION  %%%
%%%%%%%%%%%%%%%%%%%%%%%%%

\section{Introduction}
\label{Intro}

Dynamic reaction fronts formed between initially separated reactants A and B
that perform Brownian motion and react upon contact are an important
component of many physical, chemical and biological systems
\cite{Rice,HavlinBook}. Most theoretical
\cite{Galfi,Schenkel93,Haim91,CornellScaling93,LeeFront94,KozaHaim96,%
KozaJSP,KozaPHA,Sinder99,Sinder00,Sinder00Static,Magnin00}, numerical
\cite{CornellSimul} and experimental
\cite{JSPExperim91,HaimExotic,Kopel96,GelExperiment,Leger99} research has
been focused on \emph{irreversible} reactions of type A $+$ B $\to$ C, which
exhibit many unexpected phenomena. For example, the width of the reaction
zone grows with time $t$ as $t^\alpha$ with surprisingly small value of
$\alpha = 1/6$ \cite{Galfi,Schenkel93}, the center of the reaction front can
spontaneously change the direction of its motion
\cite{KozaHaim96,HaimExotic}, and the mean-field approximation of the local
reaction rate breaks down at and below the critical dimension $d_c=2$
\cite{CornellScaling93,LeeFront94,CornellSimul}.

In reality, however, most chemical reactions are reversible. The simplest
model of such a system \cite{Chopard93} is based on an assumption that
concentrations $a$, $b$, and $c$ of species A, B, and C, respectively,
effectively depend on time $t$ and only one space coordinate $x$ (even                %%%%
though the system is three-dimensional), and their evolution is governed by
three reaction-diffusion equations
\begin{eqnarray}
  \label{eq1a}
   \frac{\partial a(x,t)}{\partial t}  &=&
      D_{A} \frac{\partial^2 a(x,t)}{\partial x^2}  - R(x,t),
       \\[1ex]
  \label{eq1b}
   \frac{\partial b(x,t)}{\partial t}  &=&
      D_{B} \frac{\partial^2 b(x,t)}{\partial x^2}  - R(x,t),
       \\[1ex]
  \label{eq1c}
   \frac{\partial c(x,t)}{\partial t}  &=&
      D_{C} \frac{\partial^2 c(x,t)}{\partial x^2}  + R(x,t),
\end{eqnarray}
where the effective local reaction rate $R(x,t)$ equals to the difference
between the production (A $+$ B $\stackrel{k}{\to}$ C) and decay (C
$\stackrel{g}{\to}$ A $+$ B ) rates of species C,
\begin{equation}
   \label{eq1R}
   R(x,t) \equiv k a(x,t) b(x,t) - g c(x,t).
\end{equation}
Here  $D_{A}$, $D_{B}$, and $D_{C}$ are diffusion coefficients of species A,
B, and C, respectively, and $k,g>0$ are reaction rate constants. It is also
assumed that initially species A and B are uniformly distributed on opposite
sides of $x=0$ with concentrations $a_0$ and $b_0$, respectively,
\begin{eqnarray}
 \label{iniCond}
  a(x,0) = a_0 H(x),\quad  b(x,0) = b_0 H(-x), \quad c(x,0) = 0,
\end{eqnarray}
where $H(x)$ is the Heaviside step function (which is $0$ for $x < 0$ and $1$
for $x>0$). Such an initial condition is often adopted in experiments
\cite{JSPExperim91,HaimExotic,Kopel96,GelExperiment,Leger99} and
simplifies theoretical analysis, as it enables                 %%%%
  reduction of a three-dimensional problem to a one-dimensional one.

This model was first studied by Chopard \emph{et al.} \cite{Chopard93}. They
found that (a) the front width of a reversible reaction asymptotically
scales with time as if the process was governed solely by diffusion ($w(t)
\propto t^{1/2})$ and (b) the mean-field approximation (\ref{eq1R}) can be
safely applied for systems of spatial dimension $d=1,2,3$. However, the
fundamental problem of giving a detailed description of spatiotemporal
evolution of reversible reaction-diffusion systems remained open until quite
recently.

This problem was recently considered by  Sinder and Pelleg
\cite{Sinder99,Sinder00,Sinder00Static}. They focused their attention mainly
on the limit of a vanishingly small backward reaction rate $g$ and found
that in this limit concentrations of species A, B, and C assume the forms
typical of irreversible reactions ($g=0$) everywhere except in a very narrow
reaction zone. They confirmed the result of Ref.\ \cite{Chopard93} that
there is a crossover between intermediate-time ``irreversible'' and
large-time ``reversible'' regimes. They showed that the asymptotic reaction
rate $R$ can have one or two maxima and can even be locally negative (for
irreversible reactions $R$ always has a single maximum an can never be
negative). Moreover, they presented strong arguments supporting a conjecture
that reversible reaction-diffusion processes can be divided into two
distinct universality classes. One of them contains systems with immobile
reaction product C and asymptotically immobile reaction front, while systems
with all other combination of the control parameters form the other
universality class.

In our recent paper \cite{Koza02} we developed a new approach, enabling one
to analyze  the large-time limit of reversible reaction-diffusion systems
\emph{directly}, without having to solve the original partial differential
equations (\ref{eq1a}) -- (\ref{eq1c}) and then taking the limit
$t\to\infty$. We proved that in the large-time limit functions $a(x,t)$,
$b(x,t)$, $c(x,t)$, and $R(x,t)$ effectively depend on $x$ only through $\xi
\equiv x/\sqrt{t}$ and take on a form
\begin{eqnarray}
& a(x,t) = {\cal A}(\xi), \quad
  b(x,t) = {\cal B}(\xi), \quad
  c(x,t) = {\cal C}(\xi),        & \nonumber\\[1ex]
& R(x,t) = t^{-1}{\cal R}_1(\xi) &
\end{eqnarray}
where the scaling functions $\cal A$, $\cal B$, $\cal C$, and ${\cal R}_1$
are completely determined by four equations
\begin{eqnarray}
  \label{R0=0}
    k {\cal A}{\cal B} - g {\cal C} &=& 0, \\[1ex]
  \label{pdfA}
    D_{A} \frac{d^2 {\cal A}}{d \xi^2} +
     \frac{1}{2}\xi \frac{d {\cal A}}{d \xi}
         &=&  {\cal R}_1 \\[1ex]
  \label{pdfB}
    D_{B} \frac{d^2 {\cal B}}{\partial \xi^2} +
     \frac{1}{2} \xi \frac{d {\cal B}}{d \xi}
         &=&  {\cal R}_1 \\[1ex]
  \label{pdfC}
    D_{C} \frac{d^2 {\cal C}}{d \xi^2}
       + \frac{1}{2} \xi \frac{d {\cal C}}{d \xi}
         &=& - {\cal R}_1
\end{eqnarray}
with the boundary conditions
\begin{eqnarray}
\label{boundaryA}
  \lim_{\xi \to -\infty} {\cal A}(\xi) = a_0, \quad  \lim_{\xi \to \infty} {\cal A}(\xi) =
  0,\\
\label{boundaryB}
  \lim_{\xi \to -\infty} {\cal B}(\xi) = 0, \quad  \lim_{\xi \to \infty} {\cal B}(\xi) =
  b_0,
\end{eqnarray}
 Compared with the original problem of solving  Eqs.\
(\ref{eq1a}) -- (\ref{eq1c}), this new approach has two advantages. First,
it involves only ordinary differential equations. Second, it pertains
directly to the large-time limit.

In principle equations (\ref{R0=0}) -- (\ref{pdfC}) completely determine the
asymptotic, large-time spatiotemporal evolution of an arbitrary reversible
reaction-diffusion system. Unfortunately, they are quite complex and a
complete analytical solution is known only for the case $D_{A} = D_{B} =
D_{C}$ \cite{Koza02}. The aim of our paper is to examine these equations
numerically for other values of the control parameters.

%%%%%%%%%%%%%%%%%%%%%%%%%%%%%%%%%%%%%%%%%%%%%%%%%%%%%%%%%
%%%%%%%%%%%%%%%%%%%%    Numerics    %%%%%%%%%%%%%%%%%%%%%
%%%%%%%%%%%%%%%%%%%%%%%%%%%%%%%%%%%%%%%%%%%%%%%%%%%%%%%%%

\section{Numerical results}
\label{SectionNum}

By measuring length, time, and concentration in units of $\sqrt{D_{A}/ka_0}$,
$ 1/k a_0$, and $a_0$, respectively, the general problem of solving
(\ref{iniCond}) -- (\ref{pdfC}) for arbitrary values of $a_0$, $b_0$,
$D_{A}$, $D_{B}$, $D_{C}$, $k$, and $g$ can be reduced to the one with
\cite{Chopard93}
\begin{equation}
  \label{InitValues}
     D_{A} =1,\quad a_0 = 1, \quad k=1.
\end{equation}
We shall adopt these particular values in our further analysis. This will
leave us with four independent control parameters: $g$, $b_0$, $D_{B}$, and
$D_{C}$.

Our basic equations (\ref{R0=0}) -- (\ref{pdfC}) can be reduced to two
ordinary differential equations with two unknown functions $\cal A(\xi)$ and
${\cal B}(\xi)$,
\begin{eqnarray}
  \label{eqAC}
      \frac{
             d^2 \! \left(
                     {\cal A} + D_{C} g^{-1}{\cal A}{\cal B}
               \right)
           }{d \xi^2}
    &=&
     -\frac{1}{2} \xi
       \frac{
             d \left(
                      {\cal A} + g^{-1}{\cal A}{\cal B}
               \right)
            }{d \xi}  \\[1ex]
  \label{eqBC}
      \frac{
             d^2 \! \left(
                      D_{B}{\cal B} + D_{C} g^{-1}{\cal A}{\cal B}
                 \right)
            }{d \xi^2}
     &=&
      -\frac{1}{2} \xi
        \frac{
              d \left(
                       {\cal B} + g^{-1}{\cal A}{\cal B}
                \right)}{d \xi}
\end{eqnarray}
To solve them we employed an iterative  method. We first assumed that ${\cal
B}_0(\xi) = 0$ and, using standard techniques \cite{NR}, solved (\ref{eqAC})
as a {\em linear} ordinary differential equation for ${\cal A}_0(\xi)$ with
boundary condition (\ref{boundaryA}). We inserted this solution into
(\ref{eqBC}), which was then solved as a linear differential equation for
${\cal B}_0(\xi)$. This solution was again inserted into (\ref{eqAC}) and
used to determine the next approximation of ${\cal A}_0(\xi)$. This
procedure was repeated until a required accuracy was achieved.

Taking ${\cal B}_0(\xi) = 0$ as the first approximation leads to exact
solution for $g^{-1}=0$ (or, equivalently, $k=0$) after the first iteration
cycle, and ensures quick convergence for most choices of system parameters
except when $g \ll 1$. In this case the reaction zone is very narrow and
inside it functions ${\cal A}_0(\xi)$ and ${\cal B}_0(\xi)$ vary very
rapidly, which makes the direct iterative method unstable. This problem may
be circumvented by first solving (\ref{eqAC}) and (\ref{eqBC}) for $g \sim 1$
and then decreasing $g$ gradually until it reaches the required value, each
time employing a solution obtained for larger $g$ as an initial guess for a
smaller value of $g$. Once ${\cal A}(\xi) $ and ${\cal B}(\xi)$ have been
determined, the two remaining functions of primary interest, ${\cal C}(\xi)$
and ${\cal R}_1(\xi)$ can be calculated directly from (\ref{R0=0}) and
(\ref{pdfA}), respectively.

The iterative method cannot be applied directly for $D_{B}=D_{C}=0$, as in
this case the left-hand side of (\ref{eqBC}) vanishes and the order of this
differential equation equals 1 rather than 2. Nevertheless, since in this
very particular case $b(x,t) + c(x,t) = b_0 H(x)$, after some simple algebra
we can reduce (\ref{eqAC}) and (\ref{eqBC}) to a single equation
\begin{equation}
  \label{eqB=0C=0}
      \frac{d^2 {\cal A}(\xi)}{d \xi^2} =
     -\frac{1}{2} \xi
       \frac{d  {\cal A}(\xi)}{d \xi}
         \left(
             1+ \frac{b_0 H(\xi)}{\left[g + {\cal A(\xi)} \right]^2}
         \right)
\end{equation}
Although this equation looks very complicated, it can be solved quite easily
through standard numerical methods.

To estimate  accuracy of the iterative method, we used it to solve equations
(\ref{eqAC}) and (\ref{eqBC}) for the case of equal diffusion constants,
$D_{A} = D_{B} = D_{C} = 1$, and  compared the results with the exact
solutions obtained in Ref.\ \cite{Koza02}. For  $b_0 = 1$, $0.1$, $0.01$, $g
= 100$, $1$, $0.01$, and $-5 < \xi < 5$ we found the relative error to be
less than $10^{-6}$.

Next we used (\ref{eqAC}) -- (\ref{eqB=0C=0}) to investigate thoroughly
various combinations of system parameters. To ensure that the boundary
conditions (\ref{boundaryA}) -- (\ref{boundaryB}) are actually satisfied, we
used a rather wide range $-10 \le \xi \le 10$ and compared the results thus                %%%%%%%%%%%%%
obtained with those calculated for $-15 \le \xi \le 15$, finding no
significant differences. Moreover, upon a thorough numerical scanning of the
four-dimensional parameter space we found that the solutions of equations
(\ref{R0=0}) -- (\ref{pdfC}) are continuous functions of $D_{B}$, $D_{C}$,
$g$, and $b_0$ (even when going from one of Sinder and Pelleg's universality
classes to the other), and can be divided into three major categories
distinguished by specific forms of the local reaction rate ${\cal R}_1$.

\subsection{Reaction fronts of type I}

\begin{figure}
  \includegraphics[width=3.2in]{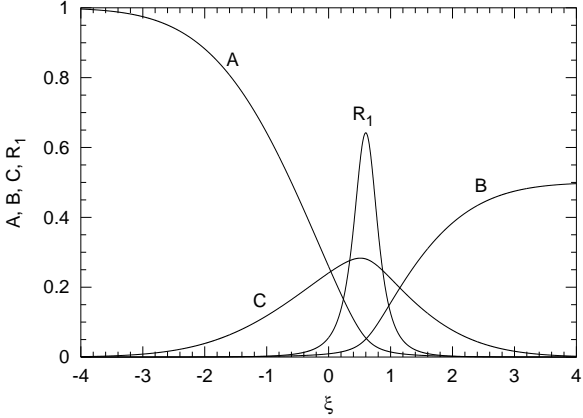}
  \caption{
    \label{Fig1}
    ${\cal A}$,
    ${\cal B}$,
    ${\cal C}$, and
    ${\cal R}_1$ (asymptotic concentrations of species A, B, C, and the scaling
    function of the reaction rate, respectively),
    as functions of $\xi \equiv x/\sqrt{t}$ for $a_0 = 1$,
    $b_0 = 0.5$, $D_{A} = D_{B} = D_{C} = 1$, $k = 1$ and $g = 0.01$. Arbitrary
    units.
 }
\end{figure}

A characteristic feature of reaction fronts of type I is that the asymptotic
reaction rate ${\cal R}_1(\xi)$ is positive for all $\xi$ and has a single
maximum, which may be identified with the reaction front center $\xi_{f}$.
 A typical example of such a reaction front is illustrated in Fig.\ \ref{Fig1},
which was obtained for $b_0 =0.5$, $D_{A} = D_{B} = D_{C} = 1$, and $g =
0.01$. As in this case $\xi_{ f} > 0$, we may say that the reaction front
moves towards the right-hand side of the system. This type of solution
always appears for $D_{A} = D_{B} = D_{C}$ \cite{Koza02} and for $g=0$
\cite{Galfi}, and so we expect it also to appear for $D_{A} \approx D_{B}
\approx D_{C} $ or for $g \ll 1$.

\begin{figure}
  \includegraphics[width=3.2in]{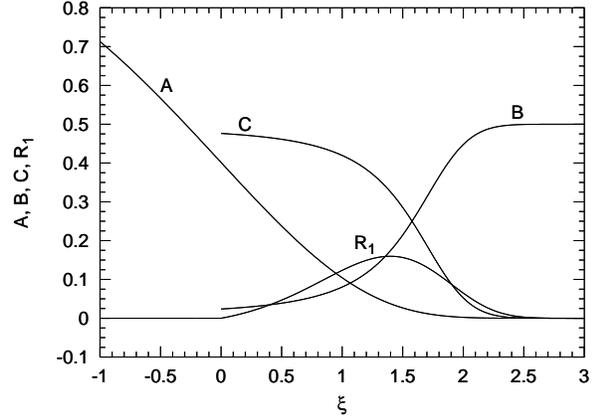}
  \caption{
    \label{Fig2}
    ${\cal A}$,
    ${\cal B}$,
    ${\cal C}$, and
    ${\cal R}_1$  as functions of $\xi \equiv x/\sqrt{t}$ for $a_0 = 1$,
    $b_0 = 0.5$, $D_{A} = 1$, $D_{B} = D_{C} = 0$, $k = 1$ and $g = 0.02$.
     Note that ${\cal B}(\xi)$ and ${\cal C}(\xi)$ vanish for $\xi < 0$.
    Arbitrary units.
 }
\end{figure}

Interestingly, it turns out that the reaction front formed in the case
$D_{B} = D_{C} = 0$  also belongs to this category. This is clearly seen in
Fig.\ \ref{Fig2}, obtained for $D_{B} = D_{C} = 0$, $b_0 = 0.5$, and $g =
0.02$. A characteristic feature of this case is discontinuity of ${\cal
B}(\xi)$ and ${\cal C}(\xi)$ at $\xi = 0$. This reflects the presence of the
Heaviside function in Eq.\ (\ref{eqB=0C=0}).

\subsection{Reaction fronts of type II}

\begin{figure}
  \includegraphics[width=3.2in]{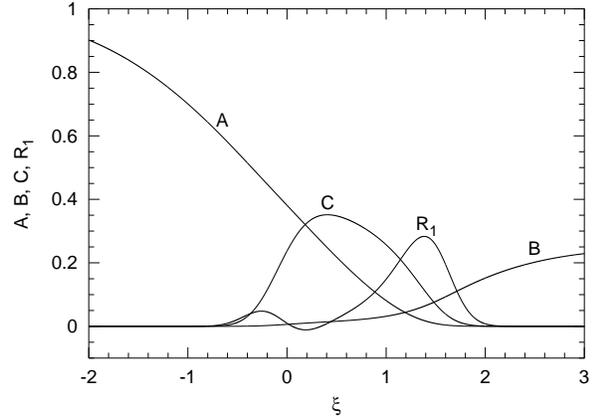}
  \caption{
    \label{Fig3}
    ${\cal A}$,
    ${\cal B}$,
    ${\cal C}$, and
    ${\cal R}_1$  as functions of $\xi \equiv x/\sqrt{t}$ for $a_0 = 1$,
    $b_0 = 0.25$, $D_{A} = D_{B} = 1$, $D_{C} = 0.01$, $k = 1$ and $g = 0.01$.
    Note that ${\cal R}_1(\xi)$ can assume negative values. Arbitrary units.
 }
\end{figure}

An example of the second type of the asymptotic solution is depicted in
Fig.~\ref{Fig3}, which was obtained for much smaller value of $D_{C} = 0.01$
(the values of other control parameters were $D_{B} = 0.5$, $b_0 =0.25$, and
$g = 0.01$). In this case ${\cal R}_1$ has two maxima $\xi_1^{\rm max}
\approx -0.26$ and $\xi_2^{\rm max} \approx 1.38$. As they are of opposite
signs, the system apparently has two reaction fronts moving at opposite
directions. Moreover, the reaction rate has one minimum, $\xi^{\rm min}
\approx 0.19$, at which it attains a negative value. In the region where
${\cal R}_1 < 0$ the backward reaction (C $\stackrel{g}{\to}$ A $+$ B) is
thus locally faster than the forward reaction (A $+$ B $\stackrel{k}{\to}$
C), although of course the global reaction rate $\int_0^\infty {\cal
R}_1(\xi)\, d\xi
> 0$.

 Formation of a
region with negative value of ${\cal R}_1$ can be understood as follows.
Consider an asymmetric reaction-diffusion system with very small diffusion
coefficient of species C ($D_{C} \ll D_{A}, D_{B}$) and a small backward
reaction rate $g$. As the reaction proceeds, the reaction front moves
through the system, leaving behind a region filled with practically immobile
and very slowly decaying reaction product C. At some moment the mobile
reaction front will leave this region, and so the backward reaction, however
small, may start to dominate. This may lead to formation of a region where
${\cal R}_1(\xi)$ attains a locally minimal, perhaps even negative value.

The dominant backward reaction should lead to production of additional
molecules of type A and B. Because $D_{A}, D_{B} \gg D_{C}$, these molecules
can easily diffuse away from the region filled with molecules C. Then, at
the other edge of the region rich in species C, they should give a
significant contribution to the forward reaction rate, forcing ${\cal
R}_1(\xi)$ to change its sign back to positive and forming the other
reaction front. Such scenario is confirmed by Fig.~\ref{Fig3}, which shows
that molecules B are present in the whole region densely occupied by
molecules C, including a region between $\xi_1^{\rm max}$ and $0$. Molecules
of type B are present in this region even though the main reaction front,
located near $\xi_2^{\rm max}$, is constantly moving away.

\begin{figure}
  \includegraphics[width=3.2in]{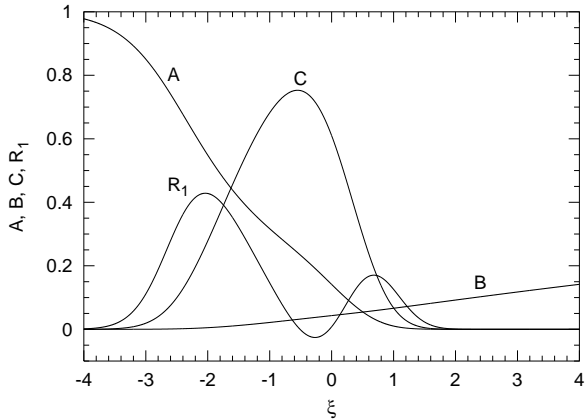}
  \caption{
    \label{Fig4}
    ${\cal A}$,
    ${\cal B}$,
    ${\cal C}$, and
    ${\cal R}_1$  as functions of $\xi \equiv x/\sqrt{t}$
    for $a_0 = 1$, $b_0 = 0.25$, $D_{A} = 1$, $D_{B} = 32$, $D_{C} = 0.01$,
    $k = 1$ and $g = 0.01$. Arbitrary units.
 }
\end{figure}

Reaction fronts of type II were first observed by Sinder and Pelleg
\cite{Sinder00Static} in systems with $D_{C} = 0$. They came to the
conclusion that for mobile reaction fronts ($x_f(t) \neq 0$) the larger
maximum is located near the point where $a(x,t) \approx b(x,t)$. However, we
found that the opposite situation is also possible. This is illustrated in
Fig.\ \ref{Fig4}, obtained for $D_{C} = 0.01$, $D_{B} = 32$, $b_0 =0.25$,
and $g = 0.01$. As we can see, in this case $a(x) \approx b(x)$ near the
second, much smaller maximum of $R_1(x)$.

We expect reaction fronts of type II to be typical of systems where either
$D_{B}$ or $D_{C}$ are much smaller than $D_{A}$. We base this conjecture on
Eqs\ (\ref{pdfB}) and (\ref{pdfC}) which ensure that if $D_{C} = 0$ or
$D_{B} = 0$ then $R_1(0) = 0$. Since $R_1(\xi)$ is continuous we may thus
expect that at least for highly asymmetric reaction fronts $R_1(\xi)$ will
attain negative values in the vicinity of $\xi = 0$.

\subsection{Reaction fronts of type III}

\begin{figure}
  \includegraphics[width=3.2in]{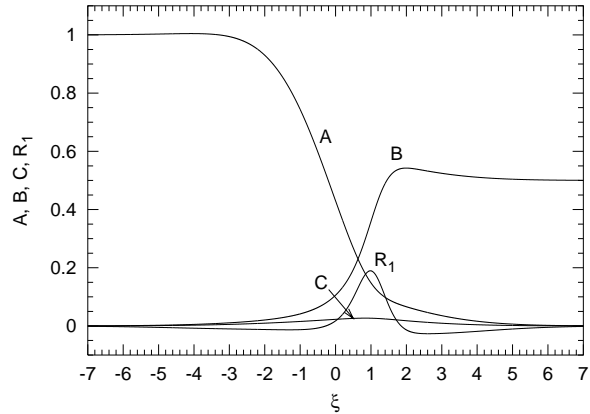}
  \caption{
    \label{Fig5}
    ${\cal A}$,
    ${\cal B}$,
    ${\cal C}$, and
    ${\cal R}_1$  as functions of $\xi \equiv x/\sqrt{t}$ for $a_0 = 1$,
    $b_0 = 0.5$, $D_{A} = 1$, $D_{B} = 0.1$, $D_{C} = 10$, $k = 1$ and $g = 2$.
    Note that ${\cal A}_0(\xi)$ and ${\cal B}_0(\xi)$
    are nonmonotonic and ${\cal R}_1$ can be negative. Arbitrary units.
 }
\end{figure}

It turns out that ${\cal A}_0(\xi)$ and ${\cal B}_0(\xi)$ may be
nonmonotonic functions of $\xi$. In this case, which we call reaction front
of type III, ${\cal R}_1(\xi)$ has a single maximum surrounded by two
minimums at which ${\cal R}_1(\xi) < 0$. All these properties are clearly
seen in Fig.\ \ref{Fig5}, obtained for $D_{C} = 10$, $D_{B} = 0.1$, $b_0
=0.5$, and $g = 2$.

This type of a reaction front can be uniquely identified by determining
whether the maximal value of $a$, denoted $a_{\rm max}$, exceeds $a_0$ (or,
similarly, whether $b_{\rm max} > b_0$). We employed this criterion in our
numerical calculations. On extensive scanning of the 4-dimensional space of
free parameters we came to the conclusion that the necessary and sufficient
condition for this type of the asymptotic reaction front reads
\begin{figure}
  \includegraphics[width=3.2in]{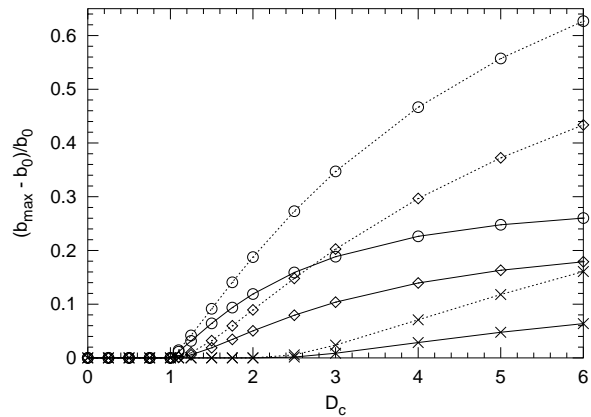}
  \caption{
    \label{Fig6}
    The maximal relative increase of the concentration of species B,
    $\tilde\Delta b_{\rm max} \equiv (b_{\rm max} - b_0)/b_0$,
    as a function of $D_C$ for $g=0.01$, $b_0 = 0.01$ (solid lines), $0.001$
    (dashed lines), $D_B = 0$ ($\circ$), 0.5 ($\diamond$), and 2 ($\times$). Arbitrary
    units.
 }
\end{figure}

\begin{equation}
   D_C > \max(D_A, D_B).
\end{equation}
However, only for $D_C \gg \max(D_A, D_B)$ is the effect really significant.
We also found that the maximal relative increase in concentrations of
species B, $\tilde\Delta b_{\rm max} \equiv (b_{\rm max} - b_0)/b_0$, is an
increasing function of $D_C$ and a decreasing function of both $b_0$ and
$D_B$. In particular  $\tilde\Delta b_{\rm max}$ turns out to be very
sensitive to changes of $b_0$, i.e.\ a parameter that can be easily
controlled experimentally.  As for $g$, $\tilde\Delta b_{\rm max}$ attains a
maximal value at $g \approx 1$ and decreases as $g \to 0 $ or $g \to
\infty$. These findings are depicted in Fig. \ref{Fig6}, which presents
$\tilde\Delta b_{\rm max}$ as a function of $D_C$ obtained for $g=0.01$,
$b_0 = 0.01$ (solid lines), 0.001 (dashed lines), and $D_B = 0$ (circles),
0.5 (diamonds), 2.0 (crosses). As we can see, $\tilde\Delta b_{\rm max}$ can
attain quite high values, exceeding 60\%.

The unusual features of this asymptotic solution can be explained as follows.
For $D_{C} \gg D_{A}, D_{B}$ molecules C quickly diffuse away from the
reaction layer. They may thus form a region where the reverse reaction
dominates the forward one, leading to ${\cal R}_1(\xi) < 0$. The same
phenomenon brings about production of additional backward reaction products A
and B outside the main reaction zone. For suitably chosen system parameters
this can result in a situation where ${\cal A}_0(\xi)$ and ${\cal B}_0(\xi)$
are nonmonotonic. This effect should become more pronounced with increased
velocity of the reaction zone (i.e., when $b_0$, $D_B \to 0$) and becomes
negligibly small as $g\to 0$ (negligible backward reaction) or $g \to \infty$
(negligible forward reaction).

%%%%%%%%%%%%%%%%%%%%%%%
%%%   CONCLUSIONS   %%%
%%%%%%%%%%%%%%%%%%%%%%%

\section{Conclusions}
\label{SectionConclusions}

We have analyzed numerically the large-time properties of reaction fronts
formed in reversible reaction-diffusion systems of type A $+$ B
$\leftrightarrow$ C. We found that, depending on the values of control
parameters, reversible reaction fronts can be divided into three categories.
In reaction fronts of type I the local reaction rate is always positive and
has a well defined, single maximum. In reaction fronts of type II the local
reaction rate has two maxima, moving in opposite directions, and a single
minimum, which can attain a negative value. In reaction fronts of type III
the local reaction rate has a single maximum surrounded by two minima, at
which it attains negative values; moreover, the concentrations of species A
and B are here locally larger than their initial values. Our numerical
calculations indicate that the condition for this type of reaction front is
given by a formula $D_C > \max(D_A, D_B)$ and that the effect of the local
increase in concentration of species A or B  can be easily controlled
experimentally through their initial concentrations $a_0$ or $b_0$.

We found that the large-time behaviour of reversible reaction-diffusion
systems is richer than that of irreversible ones. Depending on the values of
the control parameters one can expect qualitatively different asymptotic
solutions. Although the ``anomalous'' effects are rather small, we believe
that they could be observed experimentally. It would be particularly
interesting to investigate effects of nonmonotonic dependence of
concentrations of reactants A and B on the space coordinate $x$ in systems
with reaction fronts of type III. If the A $+$ B $\leftrightarrow$ C
reaction were a part of a more complex reaction scheme such that additional
reaction steps (e.g.\ precipitation) could occur only above some threshold
values of the reactant concentrations (e.g.\ nucleation thresholds), setting
$a_0$ or $b_0$ just below such a threshold value might lead to some
interesting phenomena. An example of such a complex process is formation of
the Liesegang patterns, which are quasiperiodic precipitation patterns
emerging in the wake of a mobile chemical reaction front
\cite{ChopardLiese94,ALiese01}. Our study indicates that it should be
possible to obtain similar precipitation patterns of species B in
reaction-diffusion systems with reversible reaction of type A + B
$\leftrightarrow$ C, diffusion coefficients $D_C \gg D_A, D_B$, and initial
concentrations $a_0 \gg b_0$.

\end{document}